\documentclass[aip,rsi,reprint]{revtex4-1} 
\usepackage{xcolor,graphicx}
\usepackage[utf8]{inputenc}
\usepackage{url}
\usepackage{epstopdf}

\draft 

\begin{document}

\title{Planetary gravities on a low budget: sample test of a Mars rover wheel} 

\author{G. Viera-López}
\email[]{gvieralopez@gmail.com}

\author{A. Serrano-Muñoz}

\author{J. Amigó-Vega}
\affiliation{Faculty of Telecommunications and Electronics, CUJAE, 19390 Havana, Cuba}
\affiliation{Group of Complex Systems and Statistical Physics, Physics Faculty, University of Havana, 10400 Havana, Cuba}

\author{O. Cruzata}
\affiliation{Laboratory of Laser Technology, IMRE, University of Havana, 10400 Havana Cuba}

\author{E. Altshuler}
\email[]{ealtshuler@fisica.uh.cu}
\affiliation{Group of Complex Systems and Statistical Physics, Physics Faculty, University of Havana, 10400 Havana, Cuba}

\date{\today}

\begin{abstract}
\rightskip.5in
We introduce an instrument for a wide spectrum of experiments on gravities other than our planet's. It is based on a large Atwood machine where one of the loads is a bucket equipped with a single board computer and different sensors. The computer is able to detect the falling (or rising) and then the stabilization of the effective gravity and to trigger actuators depending on the experiment. Gravities within the range 0.4 g to 1.2 g are easily achieved with acceleration noise of the order of 0.01 g. Under Martian gravity we are able to perform experiments of approximately 1.5 seconds duration. The system includes features such as WiFi and a web interface with tools for the setup, monitor and the data analysis of the experiment. We briefly show a case study in testing the performance of a model Mars rover wheel in low gravities.
\end{abstract}

\maketitle

\section{INTRODUCTION}

Setting up experiments at gravities other than Earth's is a complex physics and engineering problem. Several approaches are established, but most of them are impractical for frequent use\cite{pletser2004short}. Goldman and Umbanhowar and Altshuler \emph{et al.} developed an experimental setup based on an Atwood Machine for studying the penetration of intruders into granular media\cite{goldman2008scaling, altbucket}. More recently, Sunday \emph{et al.} have set up a facility inspired in the same principles of previous work, that has shown a good performance for effective gravities ($g_{\text{eff}}$) in the range between 0.01 g - 0.1~g\cite{sunday2016novel}.

We have upgraded our former setup for gravity-controlled experiments, originally published in 2014 for the specific case of low velocity collisions into granular matter \cite{altbucket}. We call it Lab-in-a-Bucket, a low budget solution for a wide spectrum of experiments that may require to vary the gravity. It only requires a few hundred dollars investment for hardware and electronic parts. Our instrument uses the same principle but includes different kinds of sensors and software tools that allow to customize and execute the experiments. It is also possible to track moving objects inside the capsule and to add new sensors or actuators anytime using the most common communication interfaces. We illustrate a concrete application to planetary exploration by the inclusion of a setup to test the performance of a model Mars rover wheel in low gravities.

\section{DEVICE DETAILS}

In this section we first present an overview of the system. Then the physical principles used for varying the gravity are explained. Next we describe the measures we are able to perform and the event triggers we can shoot. Finally we show other features of the system allowing us to easily customize and perform the experiments.

\begin{figure}

\includegraphics[scale=0.30]{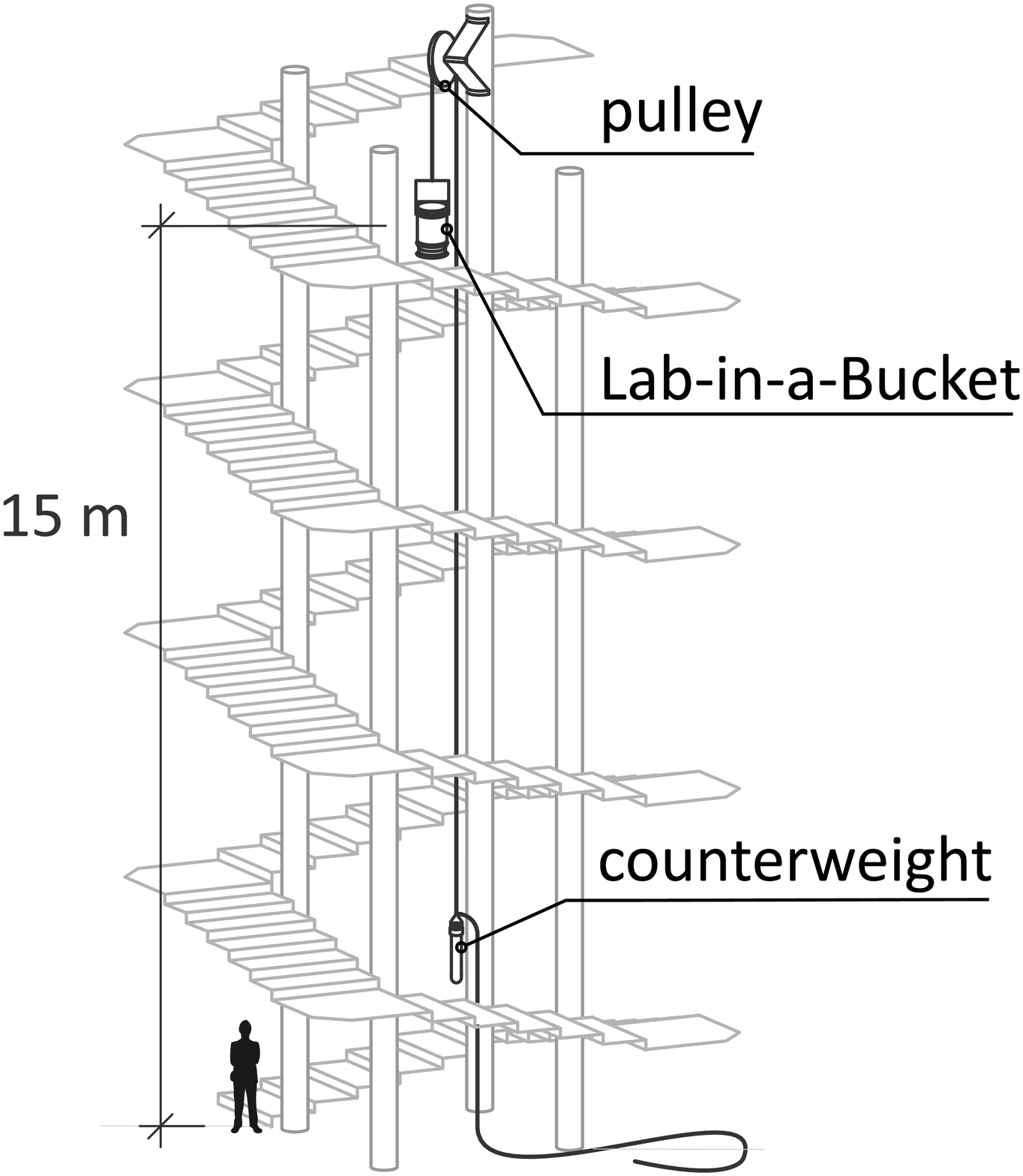}
\caption{Sketch of the system with the Lab-in-a-bucket falling as one of the loads of the Atwood machine.}
\label{fig:system}
\end{figure}

\begin{figure}
\includegraphics[scale=0.30]{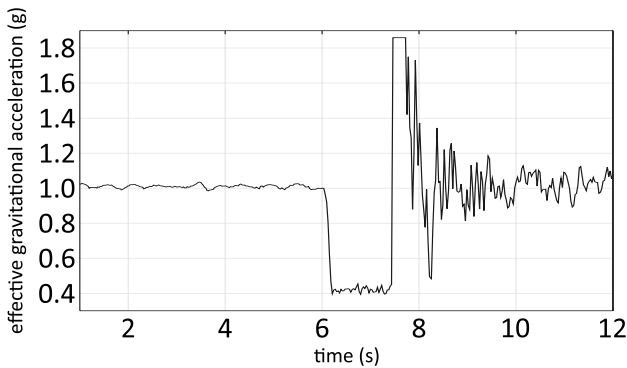}
\caption{Acceleration relative to the bucket vs time in a typical experiment. The graph shows raw data from the accelerometer, without any filtering, demonstrating the excellent stability of the system. Notice that the flat region between 6.15 s and 7.46 s approximately corresponds to the gravitational acceleration on the surface of Mars (0.4 g).}
\label{fig:measuregraph}
\end{figure}

\subsection{Overview}

The system is based on an 15-meter tall Atwood machine where one of the loads is a bucket equipped with laboratory measurement tools and the other one is the counterweight for adjusting the acceleration of the whole system, see Figure \ref{fig:system}. The instrument contains a Single Board Computer able to read data from any sensors deployed inside the bucket. The system is compatible with sensors and actuators that implement USB, I2C, SPI or UART communications protocols. The instrument creates a WiFi network to setup and monitor the experiments. We also provide software tools for real-time data processing and post-processing that are explained next.

\subsection{Mechanical principle}

We control the gravity ``felt'' inside the bucket, $g_{\text{eff}}$, by changing the mass of the counterweight following the formula:

\begin{equation}
\label{eqn:g_eff}
g_{\text{eff}} = \text{g} \left[ 1 -\frac{m_{\text{b}}-m_{\text{cw}}}{m_{\text{b}}+m_{\text{cw}}} \right]
\end{equation}

where $g \approx 9.8 m/s^2$.

The bucket can be set up to fall or rise recreating gravities lower or higher than that on Earth. We are bounded in the duration of our experiments by the height of the Atwood machine, $h$, and the desired $g_{\text{eff}}$. Equation \ref{eqn:time} gives the maximum possible duration for any desired $g_{\text{eff}}$.

\begin{equation}
\label{eqn:time}
t_{\text{max}} = \sqrt{\frac{2h}{|g - g_{\text{eff}}|}}
\end{equation}

The theoretical value expressed in \ref{eqn:time} does not take into account the time lapses spent on the stabilization of $g_{\text{eff}}$, or for stopping the moving bucket before it reaches the ground (or the pulley if it moves upwards in order to get $g_{\text{eff}}>g$). The difference in the theoretical time and the actual effective time is the price we pay for the accuracy of $g_{\text{eff}}$. Figure \ref{fig:measuregraph} shows the data of an experiment where $g_{eff}$ has been set at 0.4 g, it is possible to see that the actual effective time of the experiment is 1.31 s which is less than the maximum time (2.26 s) obtained using the equation \ref{eqn:time}. 

\subsection{Data Acquisition}

\begin{figure}
\includegraphics[scale=0.17]{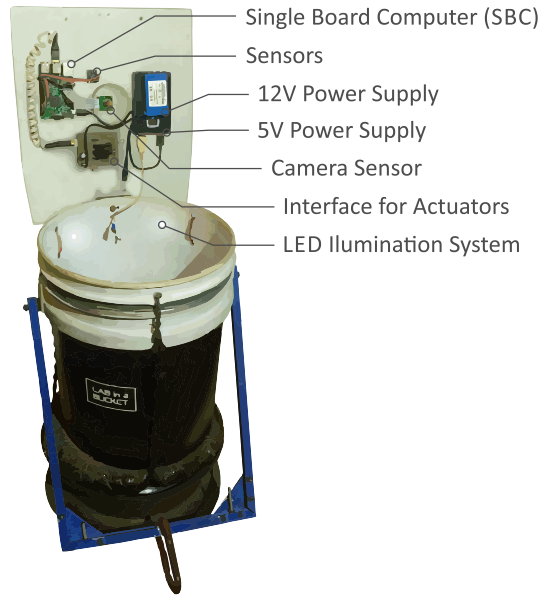}
\caption{Hardware components attached to the cover of the bucket. The cover is closed before the experiment starts.}
\label{fig:bucketzoom}
\end{figure}

We included different kinds of sensors fixed at the top of the bucket as shown in Figure \ref{fig:bucketzoom}. Three axis acceleration and gyro data are sampled using a MPU-6050 \cite{MPU6050}. Pressure and temperature inside the capsule is measured using a BMP180 \cite{BMP180} and for the magnetic field we use a HMC-5883L\cite{HMC5883L}. The computer acquires data from the sensors every 8ms and the camera sensor is sampled at 25 fps in a $800\times600$ pixel resolution. Until the experiment ends the data is temporally stored in RAM due to the writing latency imposed by the external storage devices.

As long as the experiment is taking place the information stored in RAM may be used by other software running on the computer. An example is the real-time algorithm we use to determine the stabilization of $g_{\text{eff}}$ using the measurements obtained by the accelerometer in the z-axis (see section \ref{sub:triggers} for details). Once the experiment ends all the information is stored on a SD memory plugged in the computer. 

Additional experimental information can be obtained by post-processing the data stored on the SD memory. For example, running a tracking algorithm over the camera images may return a function that describes the position of a moving object inside the bucket.

\subsection{Other features}
\label{sub:triggers}

Reacting to time events is a key feature in experiments with limited duration. We have defined two types of digital signals to optimize experimental control: Start Triggers and Stop Triggers. Launching the Start Trigger typically indicates the beginning of the experiment's effective time. It can be generated by external computers connected through the wireless network, by automatically detecting the stabilization of $g_{\text{eff}}$ or by any other convenient time event we may be able to capture. The Start Triggers can be connected to the Actuators Interface shown in Figure \ref{fig:bucketzoom}. Stop Triggers often imply the end of the experiment. It can be shot when a certain amount of time passed after the Start Trigger, but can also be generated by an external computer or by the end of the $g_{\text{eff}}$ stable period.

We designed an algorithm based on a state machine, guided by thresholds in the statistics of the z-axis acceleration signal, to automatically detect in real-time the $g_{\text{eff}}$ stabilization. Every single sample of the signal may be classified into one of the following experiment states according to the bucket motion: \emph{Stopped}, \emph{Starting}, \emph{Stable} and \emph{End}. The first sample tagged in the \emph{Stable} state (i.e., when the desired $g_{\text{eff}}$ stabilizes) may unleash the Start Trigger, the same way the first sample tagged in the \emph{End} state may unleash an Stop Trigger. 

In order to easily manage the setup of the experiments we developed a Web based Control Interface. It is possible to watch what happens inside the bucket in a real-time preview of the camera and sensors values. Every experiment gets tagged with the name, description and date. We can choose the sensors and actuators to be used and the trigger's configuration. When an experiment is completed the data obtained is stored in a file and compressed. Then the sensors data and video recorded by the camera are shown, providing a quick overview of the results. We can watch again the results of previous experiments an download the data in different formats.

The protocol of a typical experiment performed with our instrument consists in 6 steps. First, we fill the counterweight with the proper mass to get the desired $g_{\text{eff}}$. Second, we power-on and check the electronic devices inside the bucket. Third, we set the triggers and the desired configuration parameters for the experiment on the Web interface using the WiFi network. Fourth, we place both loads on the Atwood machine and hold the pulley until the oscillations stop. Fifth, we start gathering data and then release the pulley. Finally, we stop the movement of the bucket and download the sensor and video data from the web for analysis.

\section{SAMPLE APPLICATION}
\begin{figure}[!ht]
\includegraphics[scale=0.29]{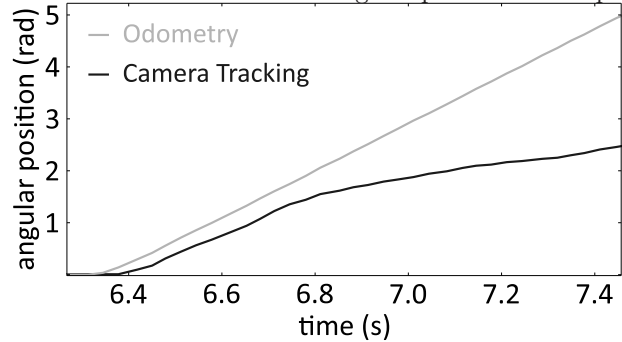}
\caption{Measurements of angular position of a rolling wheel over granular media at 0.4 g using two different methods. While the odometry quantifies the rotation of the wheel around its axis, the camera tracking measures the ``effective rotation" of the wheel based on the motion of its center of mass. Then, the graph indicates slippage on the sand surface starting slightly before 6.4 s.}
\label{grafica:slide-to-roll}
\end{figure}

Mars has been on the spotlight of the aerospace research for decades. NASA, for example, has sent rovers to study the planet. The granular nature of the planet's surface \cite{christensen2003morphology, golombek2003surface}, combined with low gravity, impose special challenges to the motion of rovers which are worth studying in detail. With our instrument it is possible to recreate an appropriate environment to study the rover's wheel performance. Inspired by the NASA rover Opportunity we 3D printed a scaled-down wheel and placed it inside our system to test its performance over Cuban sand at a gravity like that on Mars. Figure \ref{grafica:slide-to-roll} shows sample measurements made of the wheel's angular position obtained by two methods. Further results in the analysis of the performance will be published elsewhere.

\vspace{-20px}

\section{SUMMARY}  
We have presented an elastic experimental setup that allows the performance of a wide spectrum of experiments at controllable gravities, during short periods of time. The instrument shows a good performance reproducing effective gravities from 0.4 g to 1.2 g. We illustrate the capabilities of our instrument by measuring the performance of a miniature rover wheel over granular media at a gravity similar to the one found on Mars.

\bibliographystyle{apsrev}
\bibliography{rsibucket}

\begin{thebibliography}{9}
\expandafter\ifx\csname natexlab\endcsname\relax\def\natexlab#1{#1}\fi
\expandafter\ifx\csname bibnamefont\endcsname\relax
  \def\bibnamefont#1{#1}\fi
\expandafter\ifx\csname bibfnamefont\endcsname\relax
  \def\bibfnamefont#1{#1}\fi
\expandafter\ifx\csname citenamefont\endcsname\relax
  \def\citenamefont#1{#1}\fi
\expandafter\ifx\csname url\endcsname\relax
  \def\url#1{\texttt{#1}}\fi
\expandafter\ifx\csname urlprefix\endcsname\relax\def\urlprefix{URL }\fi
\providecommand{\bibinfo}[2]{#2}
\providecommand{\eprint}[2][]{\url{#2}}

\bibitem[{\citenamefont{Pletser}(2004)}]{pletser2004short}
\bibinfo{author}{\bibfnamefont{V.}~\bibnamefont{Pletser}},
  \bibinfo{journal}{Acta Astronaut.} \textbf{\bibinfo{volume}{55}},
  \bibinfo{pages}{829} (\bibinfo{year}{2004}).

\bibitem[{\citenamefont{Goldman and Umbanhowar}(2008)}]{goldman2008scaling}
\bibinfo{author}{\bibfnamefont{D.~I.} \bibnamefont{Goldman}} \bibnamefont{and}
  \bibinfo{author}{\bibfnamefont{P.}~\bibnamefont{Umbanhowar}},
  \bibinfo{journal}{Phys. Rev. E} \textbf{\bibinfo{volume}{77}},
  \bibinfo{pages}{021308} (\bibinfo{year}{2008}).

\bibitem[{\citenamefont{Altshuler et~al.}(2014)\citenamefont{Altshuler, Torres,
  González-Pita, Sánchez-Colina, Pérez-Penichet, Waitukaitis, and
  Hidalgo}}]{altbucket}
\bibinfo{author}{\bibfnamefont{E.}~\bibnamefont{Altshuler}},
  \bibinfo{author}{\bibfnamefont{H.}~\bibnamefont{Torres}},
  \bibinfo{author}{\bibfnamefont{A.}~\bibnamefont{González-Pita}},
  \bibinfo{author}{\bibfnamefont{G.}~\bibnamefont{Sánchez-Colina}},
  \bibinfo{author}{\bibfnamefont{C.}~\bibnamefont{Pérez-Penichet}},
  \bibinfo{author}{\bibfnamefont{S.}~\bibnamefont{Waitukaitis}},
  \bibnamefont{and} \bibinfo{author}{\bibfnamefont{R.~C.}
  \bibnamefont{Hidalgo}}, \bibinfo{journal}{Geophys. Res. Lett.}
  (\bibinfo{year}{2014}).

\bibitem[{\citenamefont{Sunday et~al.}(2016)\citenamefont{Sunday, Murdoch,
  Cherrier, Morales~Serrano, Valeria~Nardi, Janin, Avila~Martinez, Gourinat,
  and Mimoun}}]{sunday2016novel}
\bibinfo{author}{\bibfnamefont{C.}~\bibnamefont{Sunday}},
  \bibinfo{author}{\bibfnamefont{N.}~\bibnamefont{Murdoch}},
  \bibinfo{author}{\bibfnamefont{O.}~\bibnamefont{Cherrier}},
  \bibinfo{author}{\bibfnamefont{S.}~\bibnamefont{Morales~Serrano}},
  \bibinfo{author}{\bibfnamefont{C.}~\bibnamefont{Valeria~Nardi}},
  \bibinfo{author}{\bibfnamefont{T.}~\bibnamefont{Janin}},
  \bibinfo{author}{\bibfnamefont{I.}~\bibnamefont{Avila~Martinez}},
  \bibinfo{author}{\bibfnamefont{Y.}~\bibnamefont{Gourinat}}, \bibnamefont{and}
  \bibinfo{author}{\bibfnamefont{D.}~\bibnamefont{Mimoun}},
  \bibinfo{journal}{Rev. Sci. Instrum.} \textbf{\bibinfo{volume}{87}},
  \bibinfo{pages}{084504} (\bibinfo{year}{2016}).

\bibitem[{\citenamefont{IvenSense}(2011)}]{MPU6050}
\bibinfo{author}{\bibnamefont{IvenSense}}, \emph{\bibinfo{title}{Mpu6050}},
  \bibinfo{howpublished}{\url{http://www.daedalus.ei.tum.de/attachments/article/57/PS-MPU-6000A.pdf}}
  (\bibinfo{year}{2011}).

\bibitem[{\citenamefont{Bosch}(2013)}]{BMP180}
\bibinfo{author}{\bibnamefont{Bosch}}, \emph{\bibinfo{title}{Bmp180}},
  \bibinfo{howpublished}{\url{https://cdn-shop.adafruit.com/datasheets/BST-BMP180-DS000-09.pdf}}
  (\bibinfo{year}{2013}).

\bibitem[{\citenamefont{HoneyWell}(2010)}]{HMC5883L}
\bibinfo{author}{\bibnamefont{HoneyWell}}, \emph{\bibinfo{title}{Hmc5883l}},
  \bibinfo{howpublished}{\url{https://cdn-shop.adafruit.com/datasheets/HMC5883L_3-Axis_Digital_Compass_IC.pdf}}
  (\bibinfo{year}{2010}).

\bibitem[{\citenamefont{Christensen et~al.}(2003)\citenamefont{Christensen,
  Bandfield, Bell~III, Gorelick, Hamilton, Ivanov, Jakosky, Kieffer, Lane,
  Malin et~al.}}]{christensen2003morphology}
\bibinfo{author}{\bibfnamefont{P.~R.} \bibnamefont{Christensen}},
  \bibinfo{author}{\bibfnamefont{J.~L.} \bibnamefont{Bandfield}},
  \bibinfo{author}{\bibfnamefont{J.~F.} \bibnamefont{Bell~III}},
  \bibinfo{author}{\bibfnamefont{N.}~\bibnamefont{Gorelick}},
  \bibinfo{author}{\bibfnamefont{V.~E.} \bibnamefont{Hamilton}},
  \bibinfo{author}{\bibfnamefont{A.}~\bibnamefont{Ivanov}},
  \bibinfo{author}{\bibfnamefont{B.~M.} \bibnamefont{Jakosky}},
  \bibinfo{author}{\bibfnamefont{H.~H.} \bibnamefont{Kieffer}},
  \bibinfo{author}{\bibfnamefont{M.~D.} \bibnamefont{Lane}},
  \bibinfo{author}{\bibfnamefont{M.~C.} \bibnamefont{Malin}},
  \bibnamefont{et~al.}, \bibinfo{journal}{Science}
  \textbf{\bibinfo{volume}{300}}, \bibinfo{pages}{2056} (\bibinfo{year}{2003}).

\bibitem[{\citenamefont{Golombek}(2003)}]{golombek2003surface}
\bibinfo{author}{\bibfnamefont{M.~P.} \bibnamefont{Golombek}},
  \bibinfo{journal}{Science} \textbf{\bibinfo{volume}{300}},
  \bibinfo{pages}{2043} (\bibinfo{year}{2003}).

\end{thebibliography}

\end{document}